\documentclass[final,5p,times,twocolumn]{article}

\usepackage[utf8]{inputenc}
\usepackage{natbib}
\usepackage{graphicx}
\usepackage[margin=0.5in]{geometry}

\usepackage{lineno,hyperref}
\modulolinenumbers[5]

\begin{document}

\title{Outlier galaxy images in the Dark Energy Survey and their identification with unsupervised machine learning}
\author{Lior Shamir \\ Department of Computer Science, Kansas State University \\ 1701 Platt St \\ Manhattan, KS 66506, USA}
\date{}


\maketitle

\begin{abstract}
The Dark Energy Survey is able to collect image data of an extremely large number of extragalactic objects, and it can be reasonably assumed that many unusual objects of high scientific interest are hidden inside these data. Due to the extreme size of DES data, identifying these objects among many millions of other celestial objects is a challenging task. The problem of outlier detection is further magnified by the presence of noisy or saturated images. When the number of tested objects is extremely high, even a small rate of noise or false positives leads to a very large number of false detections, making an automatic system impractical. This study applies an automatic method for automatic detection of outlier objects in the first data release of the Dark Energy Survey. By using machine learning-based outlier detection, the algorithm is able to identify objects that are visually different from the majority of the other objects in the database. An important feature of the algorithm is that it allows to control the false-positive rate, and therefore can be used for practical outlier detection. The algorithm does not provide perfect accuracy in the detection of outlier objects, but it reduces the data substantially to allow practical outlier detection. For instance, the selection of the top 250 objects after applying the algorithm to more than $2\cdot10^6$ DES images provides a collection of uncommon galaxies. Such collection would have been extremely time-consuming to compile by using manual inspection of the data.
\end{abstract}

\section{Introduction}
\label{introduction}

The deployment of digital sky surveys driven by powerful robotic telescopes has enabled the collection of very large astronomical databases. Surveys such as Sloan Digital Sky Survey (SDSS), the Panoramic Survey Telescope and Rapid Response System (Pan-STARRS) the and the Dark Energy Survey (DES) are some of the world's most productive scientific instruments, generating databases of billions of astronomical objects. While these databases are far too large to be inspected manually, future Earth-based and space-based instruments are expected to generate even larger databases, further stressing the ability to analyze the data. Space-based instruments such as Euclid, Roman, and the Chinese Survey Space Telescope (CSST), as well as ground-based telescopes such as the Vera Rubin Observatory are expected to transform the field of astronomy by the generation of unprecedented amounts of astronomical data.

While most celestial objects in these databases can be assumed to be of known types, it is likely that these databases also contain rare objects of paramount scientific interest. The distinction between a ``peculiar'' and a ``non-peculiar'' galaxy is difficult and subjective to formalize \citep{nairn1997peculiar}. It is determined by its complex visual appearance and degree of similarity to known galaxy types. 

Galaxies that cannot be associated with a stage in a known morphological classification scheme such as the Hubble Sequence can provide unique information about the history of the Universe and galaxy evolution \citep{gillman2020peculiar}, and therefore identifying and studying of these galaxies can be of scientific value \citep{bettoni2001gas,casasola2004gas,abraham2001morphological}. Known types of peculiar galaxies are ring galaxies, which can form into a ring shape due to collisions \citep{appleton1996collisional} or instability of the galaxy bars \citep{sellwood1999formation}. Another common type of peculiar galaxies are tidally distorted galaxies, with unusual shapes formed due to the gravity field of another galaxy, leading to tidal tails or other unusual morphological features. Irregular galaxies \citep{gallagher1984structure} do not have defined expected shapes, and can contain higher amounts of gas and dust. Gravitational lenses can cause regular galaxies to appear distorted to an Earth-based observer, and while these galaxies are not peculiar they can be identified in digital sky surveys by their unusual shape. Dust lanes can also make galaxies seem unusual due to the reduction of light blocked by the dust lane \citep{mollenhoff1989peculiar,athanassoula1992existence}. Other types of peculiar galaxies include older forms such as quasar and blazars. Peculiar galaxies are of scientific importance, as they can carry substantial information about the past, present, and future Universe. For instance, strong gravitational lenses were used to determine the Hubble constant with high accuracy \citep{suyu2017h0licow,wong2020h0licow}. Statistical analysis of galaxy merger history can also be used to test the validity of cosmological models \citep{conselice2014galaxy,conselice2014evolution}. 

An example of a collection of peculiar galaxies is the Atlas of Peculiar Galaxies \citep{arp1966atlas,arp1975catalogue}. While that catalog was useful for studying peculiar galaxies, it required over a decade to prepare. Other catalogs include collections of peculiar galaxies of a certain defined type such as the collection of collisional ring galaxies \citep{madore2009atlas}. Manual observation of a large number of galaxies can lead to identification of peculiar galaxies, as was demonstrated by \cite{nair2010catalog}, who annotated a large number of galaxies and identified several peculiar galaxies through that process. \cite{kaviraj2010peculiar} identified 70 peculiar systems in stripe 82 of SDSS. Manual analysis also led to the identification of irregular and interacting galaxies imaged by the Vatican Advanced Technology Telescope \citep{taylor2005ubvr}.

Manual analysis is limited in its throughput, and therefore does not allow to analyze the extremely large databases collected by modern digital sky surveys. For instance, the Vera Rubin Observatory is expected to collect image data of more than 10$^{10}$ galaxies. Even if each galaxy can be analyzed manually in 10 seconds, analyzing the entire database will take over 3000 years of human labor. One proposed solution for increasing the throughout was to use crowdsourcing of non-expert volunteers. An example of an unusual objects identified in that manner is ``Hanny's Voorwerp" \citep{lintott2009galaxy}, as well as a large number of ring galaxies \citep{finkelman2012polar,buta2017galactic}.

As digital sky surveys become increasingly more powerful, manual identification becomes impractical. General methods to automate the analysis of galaxy images include model-driven methods such as GIM2D \citep{sim99}, GALFIT \citep{pen02}, CAS \citep{con03}, Gini \citep{abraham2003new}, Ganalyzer \citep{sha11}, and SpArcFiRe \citep{davis2014sparcfire}. Other methods are based on machine learning \citep{sha09,huertas2009robust,banerji2010,kum14,dieleman2015rotation,graham2019galaxy,mittal2019data,hosny2020classification,cecotti2020rotation,cheng2020optimizing}. For instance, \cite{huertas2009robust} used Support Vector Machines (SVM) to classify galaxies by their broad morphological type. \cite{banerji2010} demonstrated as early implementation of an artificial neural network to distinguish between elliptical and spiral galaxies. \cite{dieleman2015rotation} applied deep neural networks to estimate the expected manual annotations of certain morphological features of galaxies. More modern approaches are based on convolutional neural networks \citep{hosny2020classification,cecotti2020rotation,cheng2020optimizing}. The application of these methods to image data collected by digital sky surveys also led to catalogs \citep{huertas2015catalog,huertas2015morphologies,shamir2014automatic,kuminski2016computer,goddard2020}. Machine learning algorithms were also used to identify unusual galaxies, such as galaxy mergers \citep{margalef2020detecting}, and peculiar galaxy mergers \citep{shamir2014automatic}.

Determinstic model-driven approaches can be developed and adjusted to detect specific defined types of galaxies such as ring galaxies \citep{timmis2017catalog,shamir2020automatic} and gravitational lenses \citep{jacobs2019finding,liu2021recognition,wilde2022detecting}. Methods for automatic detection of strong lenses in large databases generated by ground-based sly surveys such as DES \citep{jacobs2019finding}, KiDS \citep{petrillo2019testing}, HSC \citep{wong2022survey}, and DECal \citep{huang2020finding}. These algorithms normally cannot match the same level of completeness as manual analysis, but their ability to scan much larger datasets allows them to identify more objects than manual detection \citep{shamir2020automatic}. Such algorithms are designed for specific and previously known morphological types, and therefore cannot identify objects of types that were not known when the algorithm was designed. That can be done by using unsupervised machine learning, where the algorithm learns automatically from the data, and can identify objects that are different from the ``typical" objects in the database. This paper describes the application of a method based on machine learning to image data acquired by the Dark Energy Survey (DES). The process leads to a collection of galaxies identified as the most different from the other ``typical'' galaxies as determined by the algorithm.

\section{Data}
\label{data}

The image data used in this study is data from the Dark Energy Survey \citep{perez2018dark,morganson2018dark,flaugher2015dark} Data Release 1 \citep{perez2018dark}. The Dark Energy Survey (DES) uses the Dark Energy Camera (DECam) of the four-meter Blanco Telescope \citep{diehl2012dark}. It covers a footprint of around $5\cdot10^3 deg^2$ \citep{abbott2018dark} in the Southern hemisphere. The primary goal of DES is the studying of dark energy, but it can also be used as a general-purpose powerful digital sky survey \citep{abbott2016dark}.

To select objects that are galaxies, the initial list of objects included objects identified as de Vaucouleurs ${r}^{1/4}$ profiles, exponential disks, or round exponential galaxies. To avoid faint objects, only objects brighter than 20.5 magnitude in one or more of the g, r or z bands were included. In DES DR1, $\sim1.9\cdot10^8$ objects met that criteria. Due to the time required to download and analyze a dataset of the size, $\sim$10\% of the data was used in this experiment, leading to a dataset of $2\cdot10^6$ objects. The images were downloaded using the {\it cutout} API of the DESI Legacy Survey, and the downloading was complete after 32 days of continuous data retrieval. The image are in the JPEG image format and dimensionality of 256$\times$256. The JPEG format does not allow accurate photometric measurements, but it allows to combine information from the g, r, and z color channels in the same image, and therefore the image contained more information about the morphology of the galaxy. The Petrosian radius of each image was used to scale the object such that the entire object fits in the image.

\section{Method}
\label{method}

Unsupervised identification of outlier images can be considered an understudied task compared to other machine vision tasks such as image classification. Unlike supervised machine learning tasks, in unsupervised machine learning the samples do not have ``ground truth'' labels, and therefore there is no training and test steps. Instead, the machine learning model attempts to identify patterns in the data without associating different patterns to different labels. In the case of automatics outlier detection, a machine learning model attempts to identify the samples that are most different from the other samples in the dataset, which needs to be done without training a system based on ground truth labels. In the case of outlier galaxies, many of the galaxies of interest may be of forms that are not yet known, and therefore no training with these galaxies is possible.

Some work on outlier galaxy detection was based on adjusting deep convolutional neural networks (DCNNs) to that task. DCNNs have demonstrated superior image classification problem, but they require a relatively large number of labeled samples for training. Rare objects often do not have a large number of instances, making it more difficult to train such model. Moreover, new objects that have never been seen before do not have any existing images, making it impossible to train a CNN model. The most common approach to apply deep neural networks to automatic identification of outlier images is by using auto-encoders, where outliers can be detected by comparing the reconstruction loss of the different images \citep{amarbayasgalan2018unsupervised,chen2018evolutionary}. Such methodology was also used to identify outlier galaxies \citep{venkat2020} or galaxy mergers \citep{margalef2020detecting}.

Deep neural networks provide good ability to analyze images, and are also relatively easy to implement by using commonly used deep learning libraries. On the other hand, due to their complex and non-intuitive nature, it is more difficult to control the noise that is part of almost all machine learning-based outlier detection systems. That is, when applying to datasets of millions of galaxies, even a small false positive rate of 1\% could make an outlier detection algorithm impractical. That requires an algorithm that can identify outlier images, but on the other hand can also reject outlier images that are the results of known non-astronomical factors. A mandatory property of such algorithm is the ability to control the trade-off between the completeness of the algorithm, and its false-positive rate. That will allow the user to sacrifice some of the outlier galaxies that will be detected in favor of limiting the false-positive rate to make the system practical in real-world settings.

The image analysis method is based on outlier detection using machine learning and a comprehensive set of numerical image content descriptors \citep{shamir2014automatic,shamir2021automatic}. In summary, the set of visual content descriptors include the entropy of the image, Radon transform \citep{lim1990two}, edge statistics, texture descriptors such as Haralick \citep{haralick1973textural}, Tamura \citep{tamura1978textural} textures, and Gabor \citep{fogel1989gabor} textures, statistics of pixel intensities, multi-scale histograms \citep{hadjidemetriou2001spatial}, Zernike polynomials \citep{teague1980image}, fractals \citep{wu1992texture}, the Gini coefficient \citep{abraham2003new}, and Chebyshev statistics. These numerical image content descriptors are described in detail in \citep{shamir2008wndchrm,shamir2010impressionism,shamir2013automatic,shamir2016morphology,schutter2015galaxy}. The source code of the method is open and publicly available \citep{shamir2017udat}. Previous studies have shown that the combination of these descriptors provide an effective numerical description of galaxy morphology \citep{sha09,schutter2015galaxy,shamir2016morphology,shamir2013automatic}.


To select numerical image content descriptors that are informative to the detection of outlier galaxies, the content descriptors are ranked by their entropy as described by Equation~\ref{entropy}.

\begin{equation}
W_f=| -1\cdot \Sigma_i P_i \cdot \log P_i  | ,
\label{entropy}
\end{equation}
where $P_i$ is the frequency of the values in the i{\it th} bin of a 10-bin histogram of the values of descriptor {\it f} measured from all images in the database. The $W_f$ weight of feature {\it f} is the entropy of that feature computed by Equation~\ref{entropy}. Low entropy of the feature reflects more consistent values, and the consistency can indicate that the values are not random, and therefore reflect the visual content.

Using the weights, the weighted distance between all pairs of images in the dataset are computed. These distance are computed by the Earth Mover's Distance (EMD). EMD is an established method for measuring distances between vectors, widely used in machine learning \citep{rubner2000earth,ruzon2001edge}. EMD can be conceptualized as an optimization problem, where the solution is the minimum work required to fill a set of holes in space with the mass of Earth. The unit of work is the work required to move an Earth unit by a distance unit. Equation~\ref{emd} shows a formal description of the EMD optimization problem.

\begin{equation}
\label{emd}
Work(X,Y,F)=\Sigma_{i=1}^n \Sigma_{j=1}^n f_{i,j}d_{i,j},
\end{equation}

where X and Y are the weighted feature vectors ${(Wx_1,x_1).....(Wx_n,x_n)}$ of size n, $f_{i,j}$ is the flow between $X_i$ and $Y_j$, and $W$ is the vector of weights. The weight vector $W$ is computed by applying Equation~\ref{entropy} to all features. The flow F is the solution of the linear programming problem: \newline \newline
$\Sigma_{i=1}^n \Sigma_{j=1}^n f_{i,j} = \min ( \Sigma_{i=1}^n Wx_i , \Sigma_{j=1}^n Wy_j ) $     \newline  \newline
With the constraints: \newline \newline
$Wx_i \geq \Sigma_{j=1}^n f_{i,j}$      \newline \newline
$Wy_j \geq \Sigma_{i=1}^n f_{i,j}$      \newline \newline

The Earth Mover’s Distance between X and Y is then defined as \newline \newline
$EMD(X,Y)=\frac{Work(X,Y,F)}{\Sigma_{i=1}^n \Sigma_{j=1}^n f_{i,j} }$  \newline

A full description of the Earth Mover's Distance method is available in \citep{rubner2000earth,ruzon2001edge}. The EMD method is effective for measuring distances between the histograms of all sets of numerical image content descriptors described in \citep{shamir2008wndchrm,shamir2010impressionism}. The distance between each pair of galaxies in the database is measured by the sum of EMD distances between all histograms.

After the similarity between each pair of images is computed, an outlier galaxy x can be detected by ranking the distances of all galaxies from galaxy x. The N{\it th} shortest distance is determined to reflect the degree of difference of the galaxies from all galaxies in the database, where $N>1$. The galaxy with the longest N{\it th} is determined to be the galaxy that is the most likely to be an outlier galaxy. The reason for using the N{\it th} shortest distance and not the shortest distance (N=1) is that in very large databases rare galaxies of the same type can appear more than once. That can lead to several galaxies similar to each other, but different from all other galaxies in the database. A short distance between the two galaxies might therefore reflect two or more outlier galaxies that are similar to each other but could be different from all other galaxies. Therefore, using the shortest distance as a measurement of how different a galaxy is from all other galaxies might lead the algorithm to a high number of false negatives.

On the other hand, some images might be different from other galaxy images in the database for certain non-astronomical reasons such as the impact of nearby very bright stars on the imaging. Examples of such images are provided in Section~\ref{results}. Such artefacts make the images look very different than regular galaxies, but they are also not rare. Therefore, taking the N{\it th} distance can allow to avoid some of these artefacts that are common in the database. If the artefact is not common, the algorithm might falsely flag it as an outlier, increasing the false positive rate of the algorithm. In any case, the large databases of current and future astronomical sky surveys require the ability to handle the trade-off between completeness and false positive rate, as even a small false positive rate might make such algorithm impractical due to the very large number of non-outlier galaxies identified.

\section{Results}
\label{results}

The outlier detection method described in Section~\ref{method} was applied to the image data collected by the Dark Energy Survey as described in Section~\ref{data}. Due to the large number of galaxies, the $2\cdot10^6$ galaxies were separated into 100 sets of $2\cdot10^4$ galaxies. Then, the algorithm described in Section~\ref{method} was applied to each of the 100 sets, returning the top 30 most peculiar images as determined by the algorithm. The value of N was set to 50. That led to a dataset of 3,000 galaxies that could be considered as possible peculiar galaxies.

As also mentioned in Section~\ref{method}, the algorithm is not expected to be fully accurate in the identification of outlier galaxies. Many of the galaxies identified as outlier galaxies are not expected to indeed be of scientific interest, and therefore manual selection is required. The advantage of using the algorithm is that the manual selection is applied to 3,000 galaxies, which is several orders of magnitude less than the initial set of $2\cdot10^6$ DES galaxies. The 3,000 galaxies picked by the algorithms can be separated into regular galaxies, artefacts, or true positives of galaxies that could be of scientific interest.

\begin{figure}[ht]
\centering
\includegraphics[scale=0.24]{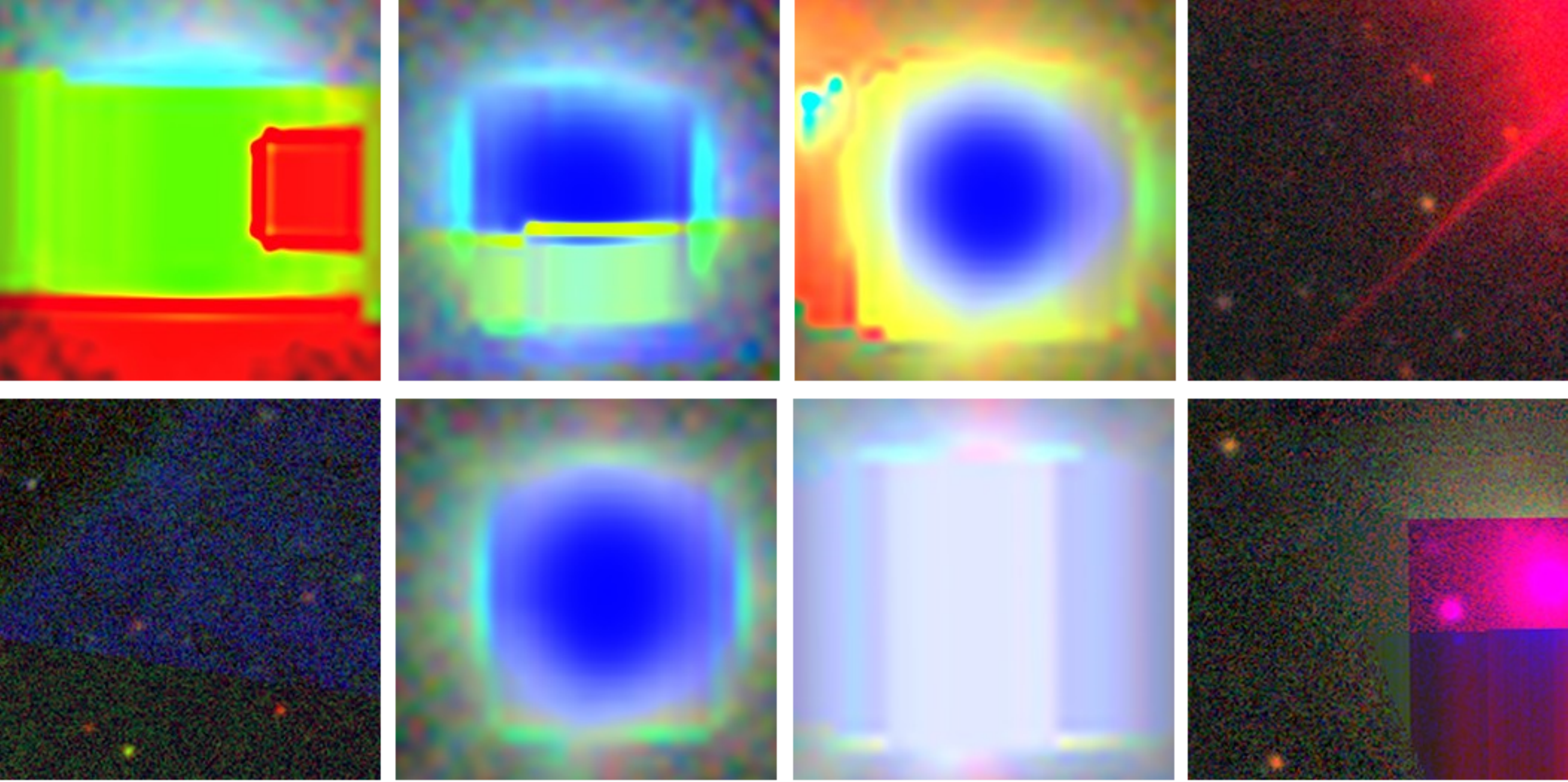}
\caption{Objects considered galaxies that were detected by the algorithm as outliers, but visual inspection shows that these objects are not of astronomical interest.}
\label{artefacts}
\end{figure}

\begin{figure}[ht]
\centering
\includegraphics[scale=0.24]{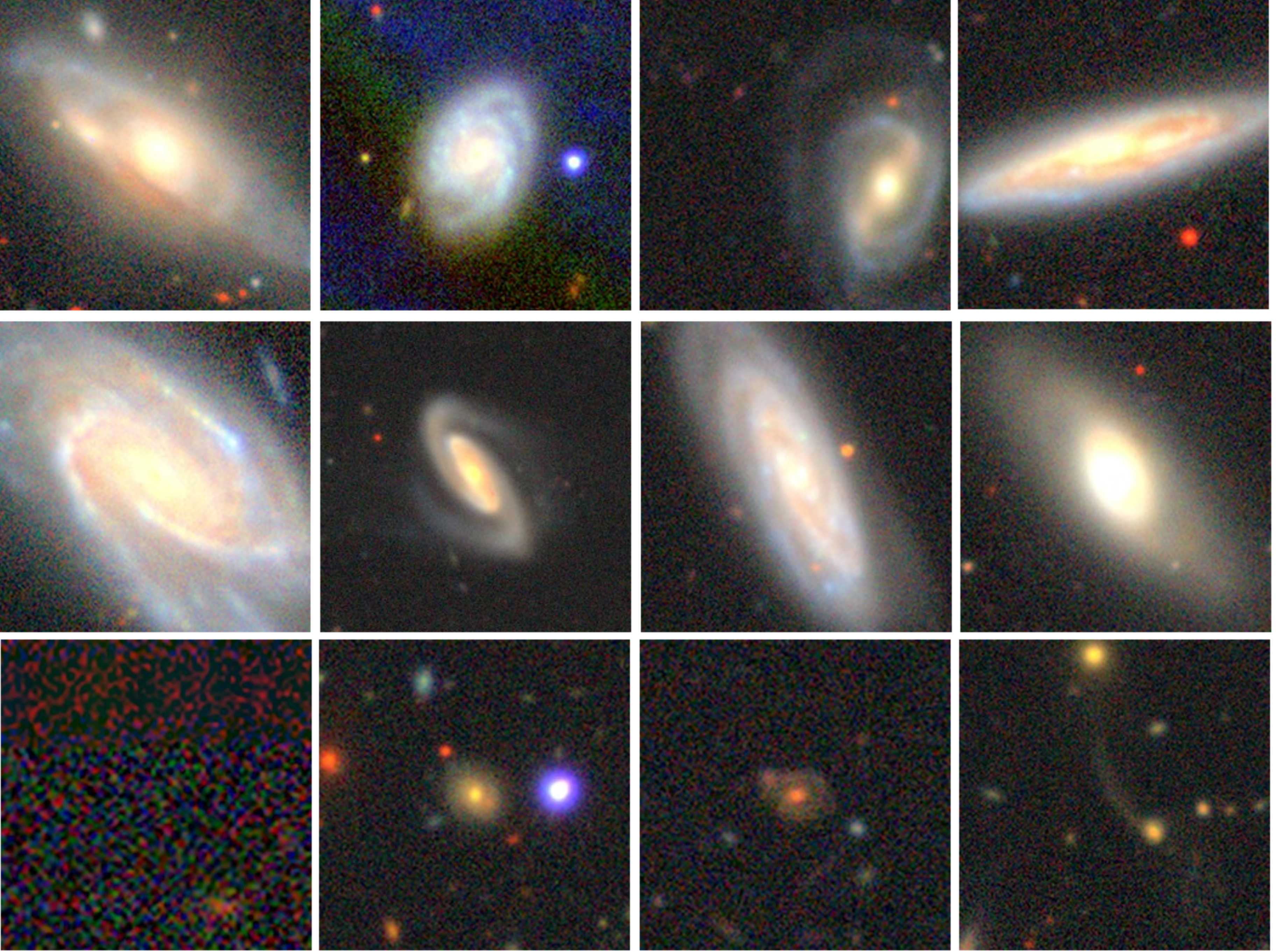}
\caption{Galaxies that were detected by the algorithm as outliers, but seem to be regular galaxies by visual inspection.}
\label{regular}
\end{figure}

Figure~\ref{artefacts} shows examples of images identified by the algorithm that are different from a regular galaxy, but the difference cannot be considered of particular astronomical interest. As the figure shows, these outlier images are different from most other galaxy images, although the reasons for the differences are not necessarily of astronomical origin. The identification of artefacts and unusual images driven by non-astronomical reasons is based on previous knowledge, as these forms of outlier images are relatively common. It is therefore theoretically possible that true outlier galaxies that seem similar to common artefacts might not be identified. Figure~\ref{regular} shows galaxies identified by the algorithm as outliers, although visual inspection shows that the galaxies do not have unusual features.

The definition of peculiar or unusual galaxies is not necessarily formal \citep{nairn1997peculiar}, making the manual identification of all peculiar galaxies a task that is not considered of high precision. Some galaxies of scientific interest might therefore not be identified as peculiar galaxies. In this study, the selection of peculiar galaxies was done such that galaxies that belong in a known type of usual galaxies were selected based on their visual appearance. That selection was done based on previous knowledge of these galaxies, although many of the galaxies that were detected are of morphology that does not necessarily have an existing known similar instance. Like with the removal of artefact, that process of manual selection can also lead to the loss of some galaxies of scientific interest. On the other hand, identification of peculiar galaxies when done purely by manual labor can also lead to incompleteness of the output. A notable example is the ``Hanny's voorwerp'' galaxy \citep{lintott2009galaxy}, which was annotated as a regular galaxy by two more than dozens different observers, until it was identified as an unusual galaxy of scientific interest.

\begin{figure}[ht]
\centering
\includegraphics[scale=0.37]{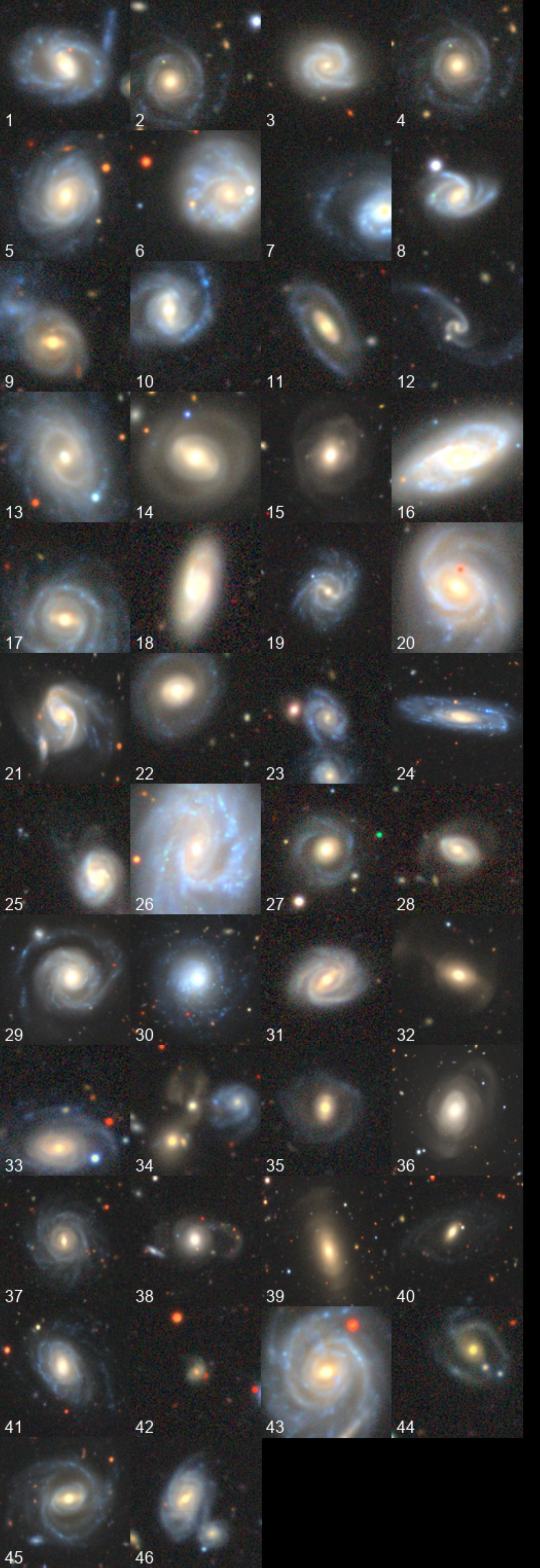}
\caption{Galaxies with detached segments that were detected by the algorithm.}
\label{detached_segments_fig}
\end{figure}

\begin{figure}[ht]
\centering
\includegraphics[scale=0.37]{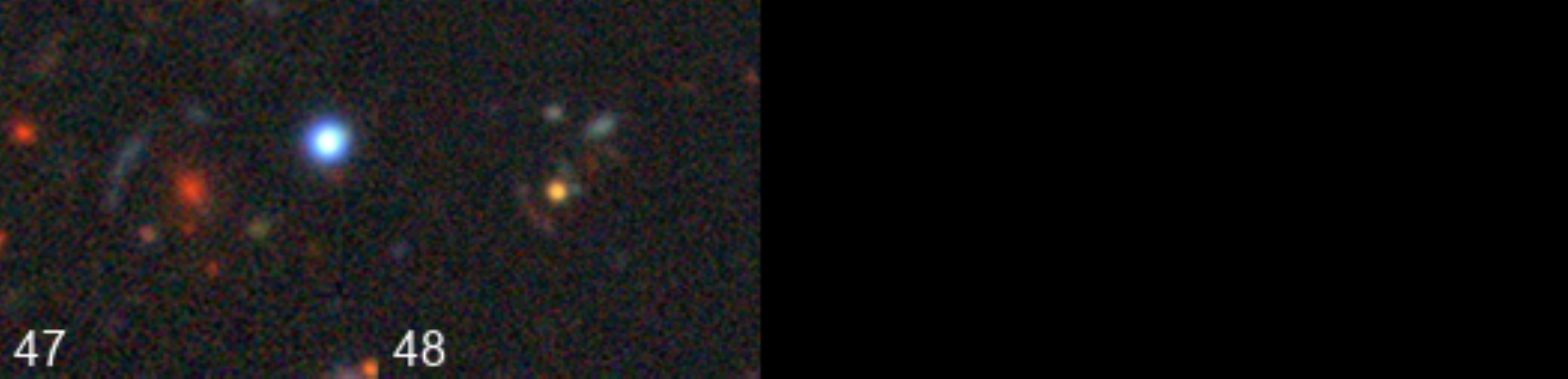}
\caption{Galaxies that are possible gravitational lenses.}
\label{gravitational_lenses_fig}
\end{figure}

\begin{figure}[ht]
\centering
\includegraphics[scale=0.35]{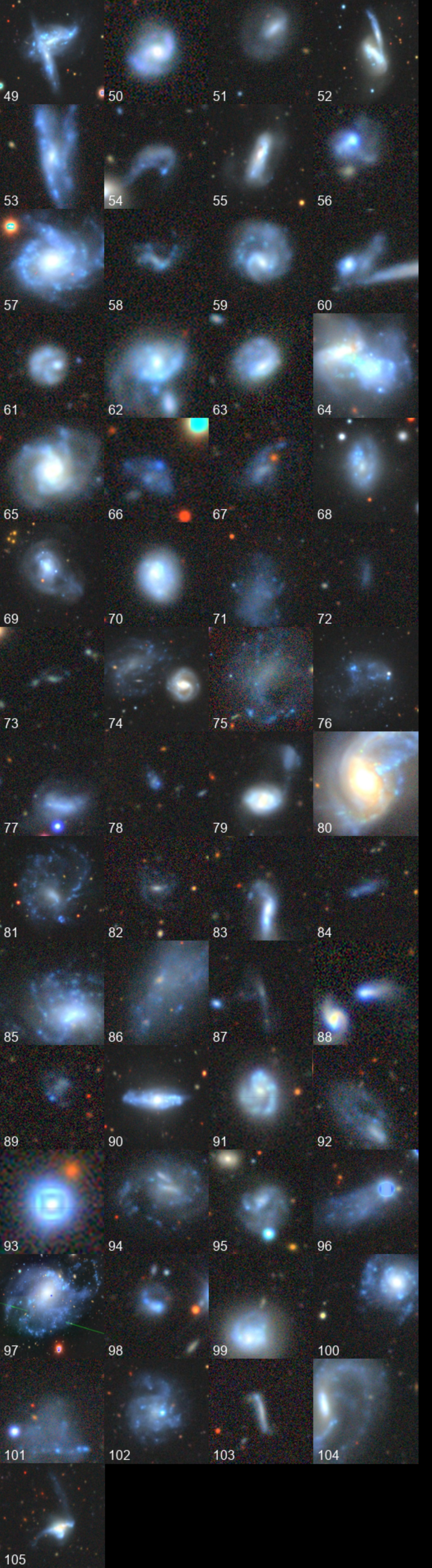}
\caption{Irregular blue galaxies detected in DES.}
\label{irregular_blue_fig}
\end{figure}

\begin{figure}[ht]
\centering
\includegraphics[scale=0.32]{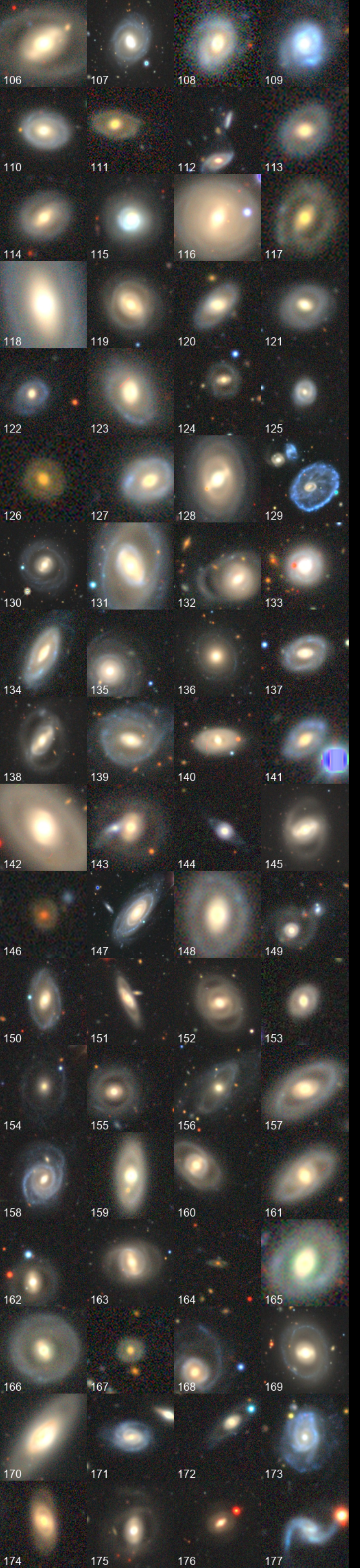}
\caption{Ring galaxies detected in DES.}
\label{ring_fig}
\end{figure}

\begin{figure}[ht]
\centering
\includegraphics[scale=0.37]{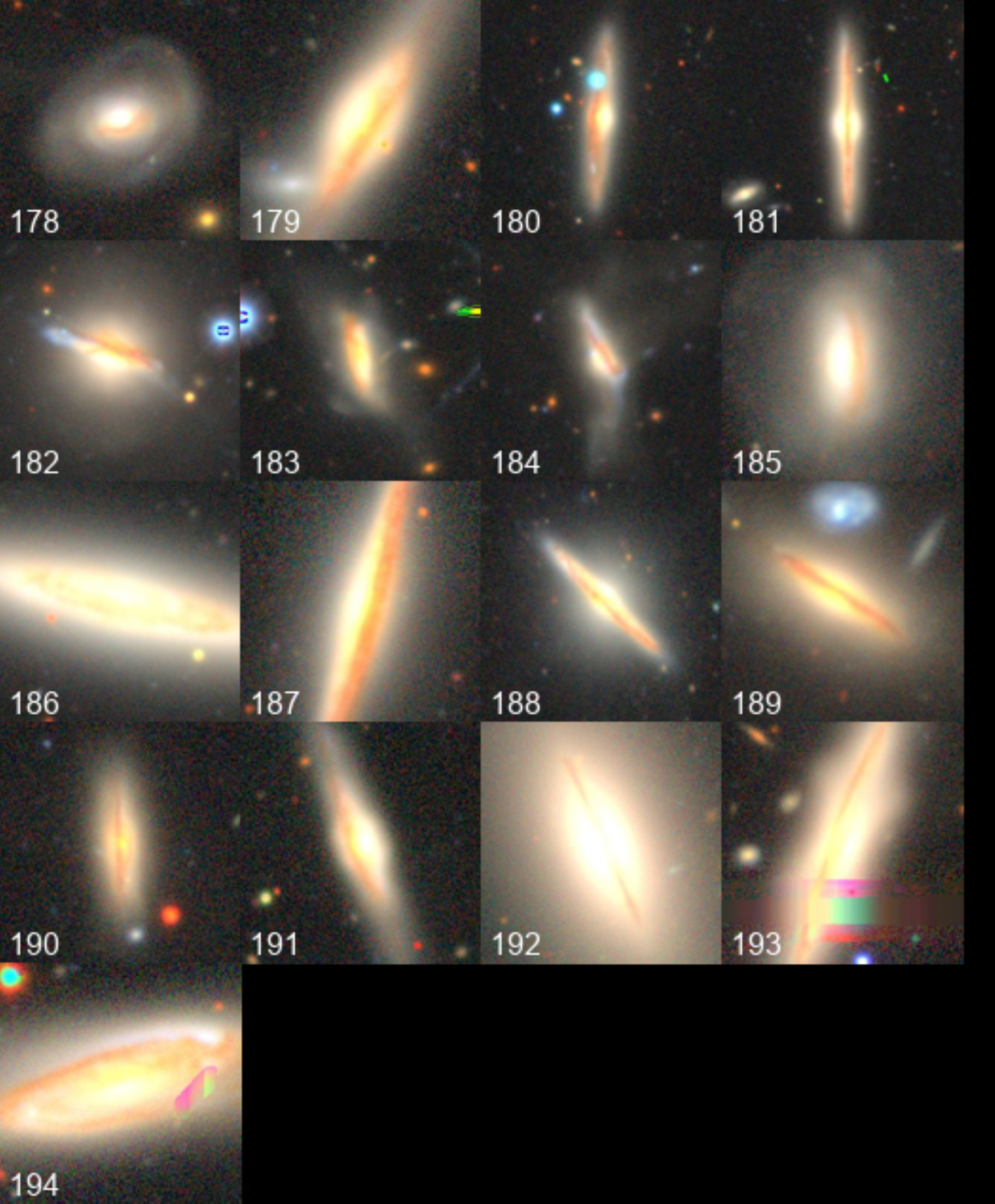}
\caption{Galaxies with dust lanes.}
\label{dust_lane_fig}
\end{figure}

\begin{figure}[ht]
\centering
\includegraphics[scale=0.4]{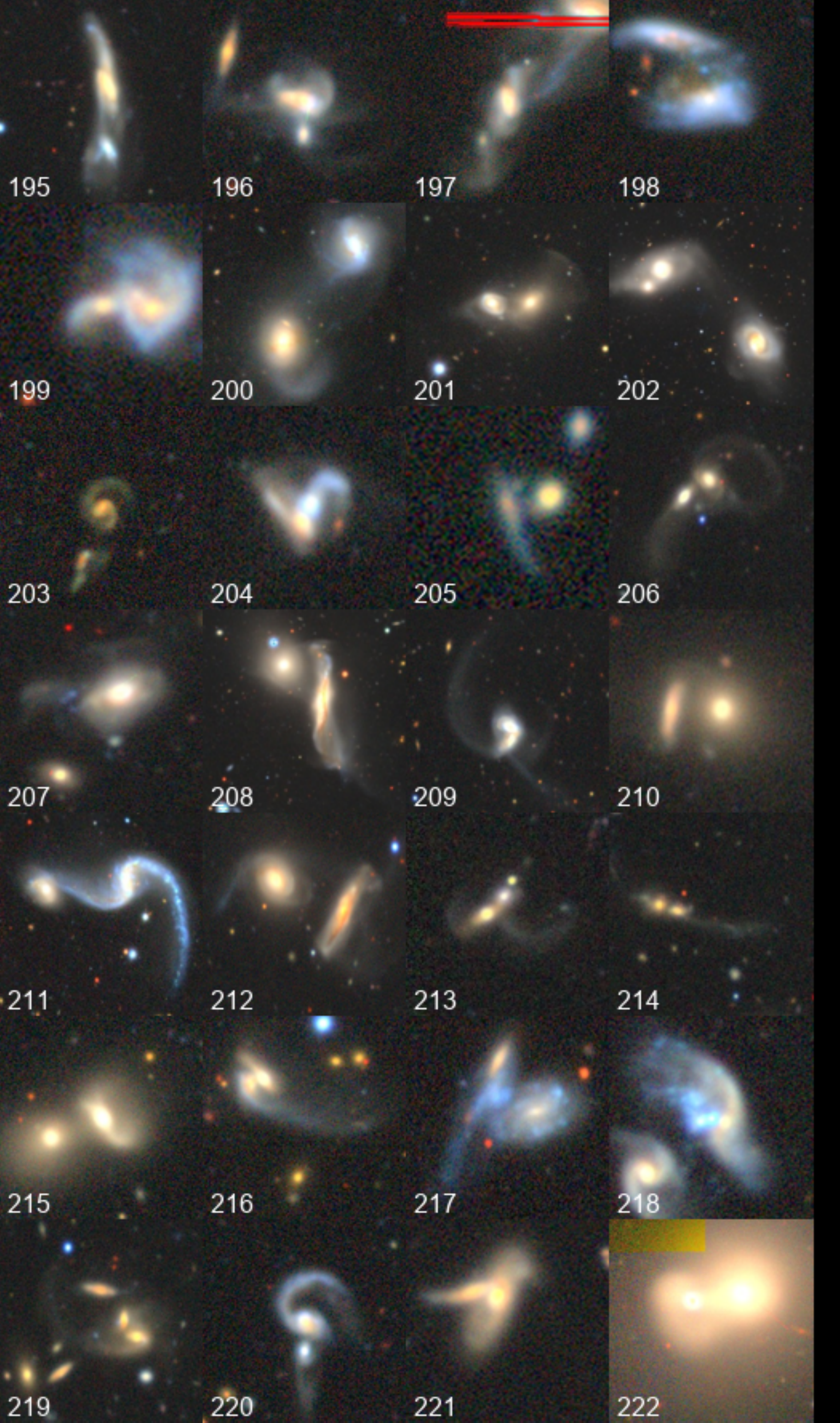}
\caption{Images of detected objects that can be tidally distorted systems.}
\label{tidially_distorted_fig}
\end{figure}

\begin{figure}[ht]
\centering
\includegraphics[scale=0.4]{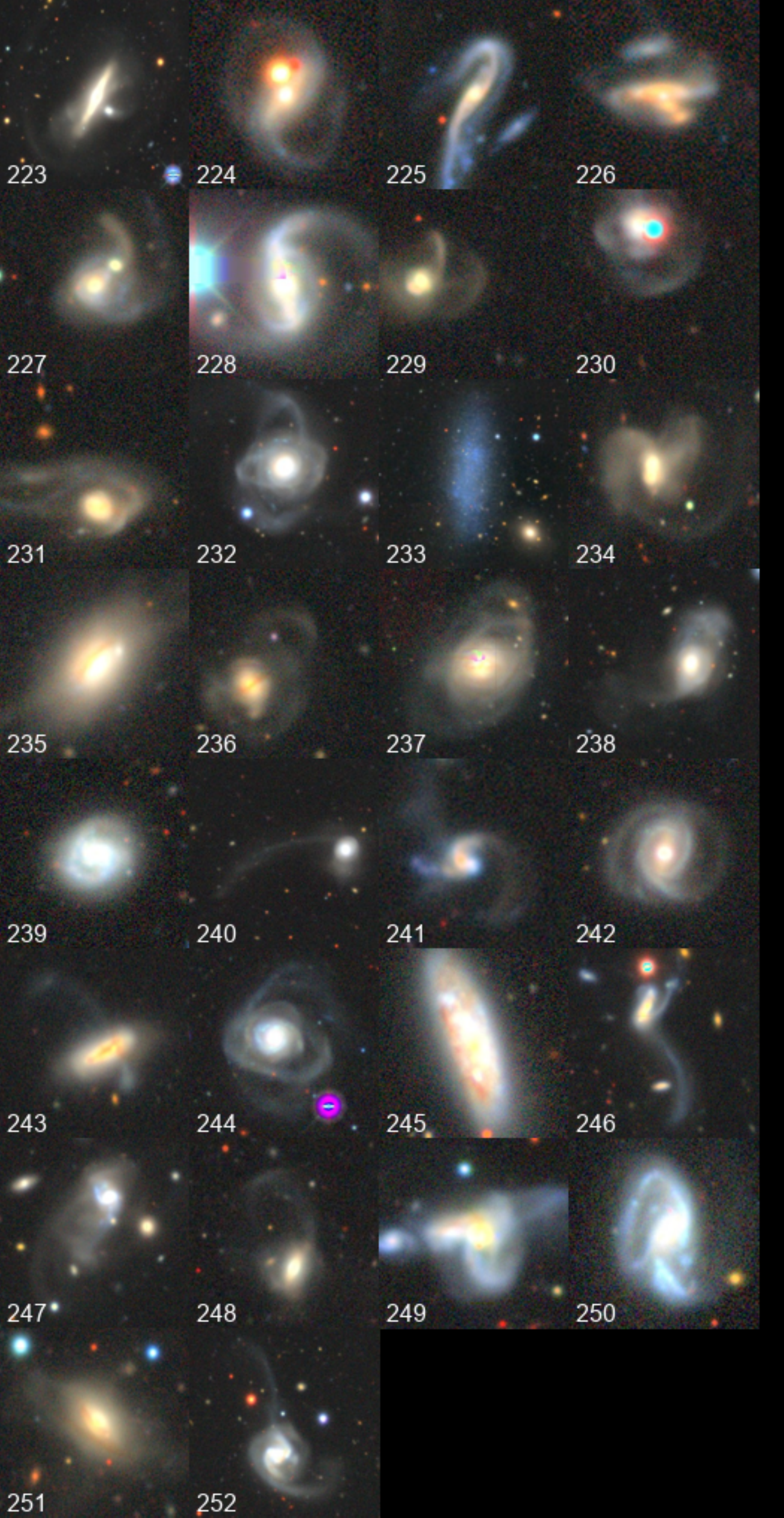}
\caption{Unusual galaxies that are not associated to the previous categories.}
\label{others_fig}
\end{figure}

From the galaxies identified by the algorithm, 250 galaxies were identified by visual inspection as galaxies that could have certain features that make these galaxies different from most other galaxies. These galaxies were separated into several categories. Tables~\ref{detached_segments} through~\ref{others} show the equatorial coordinates of the galaxies detected by the algorithm, and separated into the different categories. The images of the galaxies are displayed in Figures~\ref{detached_segments_fig} through~\ref{others_fig}.

Tables~\ref{detached_segments} and~\ref{dust_lane} list galaxies with detached segments and dust lanes. Such galaxies are not necessarily considered peculiar galaxies, but most galaxies do not have clear large detached segments or dust lanes. Object 129 is the ``Cartwheel" galaxy. While that object is known, its detection shows that the algorithm can detect unusual objects automatically. Table~\ref{tidially_distorted} shows gravtiationally interacting systems such that the interactions change the shape of at least one of the galaxies in the system. Figure~\ref{tidially_distorted_fig} displays the images of the galaxies in that table. Such galaxies are common in the Atlas of Peculiar Galaxies \citep{arp1966atlas}, but since these system are relatively rare their identification by manual inspection is a labor consuming task.

Table~\ref{gravitational_lenses} shows possible gravitational lenses. These objects are not included in known previous catalogs of gravitational lenses such as the CASTLES survey of gravitational lenses \citep{kochanek1999results}, the catalog of SDSS gravitational lens candidates \citep{inada2012sloan}, gravitational lenses detected in COSMOS \citep{faure2008first}, or a survey powered by a group finding algorithm \citep{wilson2016spectroscopic}. These lenses are also not present in detected gravitational lenses in HSC \citep{wong2018survey}, or in catalogs compiled by using convolutional neural networks applied to DES \citep{jacobs2019extended}, or the VST Optical Imaging of the CDFS and ES1 survey \citep{gentile2022lenses}. 

Table~\ref{others} list objects that cannot be associated clearly with any of the groups, and these galaxies are shown in Figure~\ref{others_fig}. For instance, object 252 is a galaxy with two dense arms, seemingly embedded in another, less dense, structure. The top of the system features another sparse and long arm that is not necessarily aligned with the other arms of the galaxy. Object 228 features a ring as well as a one spiral arm. Objects 232 and 244 has several rings. Object 250 also has a ring, but also has several other features making it more difficult to characterize the galaxy as a ring galaxy. Although these shapes can be the results of gravitational interaction between two or more objects, the images do not show another object that can lead to the peculiar shapes. Fully understanding each of these systems might require further detailed observation of these systems.

\begin{table*}[h!t]
{
\centering
\begin{tabular}{|l|c|c|c|c|c|c|c|c|c|c|}
\hline
ID & RA & Dec  & & ID & RA & Dec & & ID & RA & Dec \\
\hline
1 & 25.0160 & -43.423 & & 2 & 53.3290 & -39.219 & & 3 & 53.1833 & -39.151 \\
4 & 53.3321 & -39.221 & & 5 & 29.5303 & -39.041 & & 6 & 93.2225 & -38.770 \\
7 & 40.0661 & -38.481 & & 8 & 66.9333 & -37.460 & & 9 & 18.2232 & -34.734 \\
10 & 19.3597 & -34.737 & & 11 & 48.0650 & -34.784 & & 12 & 41.6079 & -33.938 \\
13 & 92.2672 & -33.710 & & 14 & 82.3580 & -29.968 & & 15 & 11.5389 & -24.646 \\
16 & 13.8953 & -24.153 & & 17 & 18.2169 & -23.816 & & 18 & 14.9967 & -22.885 \\
19 & 14.2179 & -22.104 & & 20 & 44.0267 & -14.186 & & 21 & 15.0158 & -4.9402 \\
22 & 26.1296 & -4.9703 & & 23 & 12.5307 & -4.4804 & & 24 & 10.9840 & -4.2419 \\
25 & 12.1810 & -4.2121 & & 26 & 26.3561 & -3.8271 & & 27 & 9.93519 & -65.011 \\
28 & 9.53072 & -65.120 & & 29 & 12.5187 & -3.5511 & & 30 & 15.7978 & -3.6055 \\
31 & 1.42831 & -3.0761 & & 32 & 4.27571 & -2.6985 & & 33 & 8.59889 & -2.8351 \\
34 & 1.85732 & -2.3476 & & 35 & 9.0261 & -2.256. & & 36 & 13.0401 & -2.2339 \\
37 & 0.55003 & -1.8722 & & 38 & 3.04020 & -1.8381 & & 39 & 34.1008 & -64.027 \\
40 & 31.1652 & -60.958 & & 41 & 37.8805 & -60.589 & & 42 & 30.8068 & -60.123 \\
43 & 23.0091 & -53.861 & & 44 & 25.0463 & -49.265 & & 45 & 17.2175 & -47.142 \\
46 & 13.5656 & -3.5828 & &  & &    & &  & &  \\
\hline
\end{tabular}
\caption{The equatorial coordinates of detected objects with detached segments.}
\label{detached_segments}
}
\end{table*}

\begin{table*}[ht]
{
\centering
\begin{tabular}{|l|c|c|c|c|c|c|}
\hline
ID & RA & Dec  & & ID & RA & Dec  \\
\hline
47 & 36.0262 & -61.719 & & 48 & 35.8424 & -60.673 \\
\hline
\end{tabular}
\caption{The equatorial coordinates of objects detected as possible gravitational lenses.}
\label{gravitational_lenses}
}
\end{table*}

\begin{table*}[h!t]
{
\centering
\begin{tabular}{|l|c|c|c|c|c|c|c|c|c|c|}
\hline
ID & RA & Dec  & & ID & RA & Dec & & ID & RA & Dec \\
\hline
49 & 19.9835 & -41.233 & & 50 & 9.23238 & -39.989 & & 51 & 24.5640 & -39.939 \\
52 & 88.7023 & -39.764 & & 53 & 70.2373 & -39.496 & & 54 & 23.4841 & -39.085 \\
55 & 26.2304 & -38.737 & & 56 & 38.8150 & -36.994 & & 57 & 57.0292 & -36.701 \\
58 & 64.5721 & -36.764 & & 59 & 48.0095 & -36.426 & & 60 & 37.6970 & -36.315 \\
61 & 73.0834 & -35.645 & & 62 & 5.89178 & -34.969 & & 63 & 12.1449 & -34.899 \\
64 & 90.0192 & -33.919 & & 65 & 6.13306 & -33.087 & & 66 & 26.5894 & -29.549 \\
67 & 10.6990 & -24.896 & & 68 & 13.8724 & -24.161 & & 69 & 11.8880 & -22.804 \\
70 & 11.8071 & -21.708 & & 71 & 10.9322 & -21.377 & & 72 & 0.74651 & -4.8807 \\
73 & 13.3881 & -4.6834 & & 74 & 15.8261 & -4.6680 & & 75 & 8.24426 & -65.300 \\
76 & 15.5969 & -4.4105 & & 77 & 23.2759 & -4.5320 & & 78 & 10.3841 & -3.8822 \\
79 & 16.7885 & -4.0034 & & 80 & 27.8298 & -4.0559 & & 81 & 26.6577 & -3.7783 \\
82 & 2.08945 & -3.6142 & & 83 & 13.5512 & -3.4408 & & 84 & 9.21452 & -2.7584 \\
85 & 2.25343 & -2.3795 & & 86 & 1.23298 & -64.769 & & 87 & 7.49207 & -2.3000 \\
88 & 0.76150 & -2.0754 & & 89 & 0.84586 & -1.4346 & & 90 & 15.7047 & -1.4702 \\
91 & 12.1583 & -64.520 & & 92 & 12.8725 & -64.356 & & 93 & 16.8820 & -64.233 \\
94 & 23.6451 & -64.353 & & 95 & 14.1570 & 0.10382 & & 96 & 25.8844 & -63.967 \\
97 & 18.2076 & 0.98033 & & 98 & 8.25643 & -62.567 & & 99 & 12.5641 & -59.892 \\
100 & 12.8893 & -59.708 & & 101 & 12.8893 & -59.080 & & 102 & 16.3372 & -57.229 \\
103 & 86.6126 & -53.999 & & 104 & 18.3707 & -52.304 & & 105 & 10.1921 & -51.463 \\
\hline
\end{tabular}
\caption{The coordinates of detected irregular blue galaxies.}
\label{irregular_blue}
}
\end{table*}

\begin{table*}[h!t]
{
\centering
\begin{tabular}{|l|c|c|c|c|c|c|c|c|c|c|}
\hline
ID & RA & Dec  & & ID & RA & Dec & & ID & RA & Dec \\
\hline
106 & 21.8298 & -44.064 & & 107 & 14.2471 & -43.842 & & 108 & 316.708 & -40.356 \\
109 & 22.3453 & -39.905 & & 110 & 29.5910 & -39.879 & & 111 & 20.3687 & -39.677 \\
112 & 21.8018 & -39.836 & & 113 & 22.1732 & -39.637 & & 114 & 24.1116 & -39.413 \\
115 & 29.4474 & -39.440 & & 116 & 22.1516 & -39.307 & & 117 & 18.0015 & -38.913 \\
118 & 21.4559 & -39.105 & & 119 & 21.9585 & -39.030 & & 120 & 43.6317 & -39.067 \\
121 & 75.2043 & -38.876 & & 122 & 24.6654 & -38.764 & & 123 & 80.6596 & -37.752 \\
124 & 88.6841 & -36.522 & & 125 & 31.7459 & -35.719 & & 126 & 44.6345 & -34.677 \\
127 & 20.9356 & -34.461 & & 128 & 41.7983 & -34.427 & & 129 & 9.43734 & -33.705 \\
130 & 18.7599 & -32.234 & & 131 & 18.0802 & -32.061 & & 132 & 20.2459 & -29.211 \\
133 & 10.4466 & -24.387 & & 134 & 12.7404 & -23.557 & & 135 & 10.3802 & -22.641 \\
136 & 15.3996 & -22.059 & & 137 & 48.4839 & -14.969 & & 138 & 25.6132 & -14.533 \\
139 & 20.3018 & -14.243 & & 140 & 20.3157 & -13.793 & & 141 & 23.3719 & -13.863 \\
142 & 6.83340 & -4.8914 & & 143 & 10.3298 & -4.9730 & & 144 & 12.3963 & -4.9378 \\
145 & 14.4237 & -4.9554 & & 146 & 16.1468 & -4.6820 & & 147 & 29.1036 & -4.4724 \\
148 & 6.31877 & -2.7616 & & 149 & 28.2879 & -64.772 & & 150 & 34.7829 & -64.804 \\
151 & 9.02329 & -2.3321 & & 152 & 9.51328 & -2.3177 & & 153 & 1.38100 & -2.1122 \\
154 & 0.04351 & -1.5109 & & 155 & 27.4904 & -64.487 & & 156 & 23.7545 & -64.243 \\
157 & 26.0110 & -64.370 & & 158 & 18.3670 & 0.14845 & & 159 & 14.0249 & 0.62863 \\
160 & 26.7319 & -64.019 & & 161 & 9.87239 & -63.768 & & 162 & 11.2942 & -63.728 \\
163 & 36.1261 & -61.662 & & 164 & 32.1129 & -60.141 & & 165 & 233.126 & 52.6243 \\
166 & 230.236 & 54.6749 & & 167 & 57.6141 & -54.149 & & 168 & 24.4276 & -53.779 \\
169 & 19.8673 & -51.772 & & 170 & 24.7202 & -49.785 & & 171 & 26.0838 & -49.760 \\
172 & 23.2804 & -49.199 & & 173 & 22.6231 & -48.838 & & 174 & 357.852 & 10.8401 \\
175 & 13.0401 & -2.2339 & & 176 & 358.825 & 11.2425 & & 177 & 359.842 & 10.6572 \\
\hline
\end{tabular}
\caption{The coordinates of detected possible ring galaxies.}
\label{rings}
}
\end{table*}

\begin{table*}[h!t]
{
\centering
\begin{tabular}{|l|c|c|c|c|c|c|c|c|c|c|}
\hline
ID & RA & Dec  & & ID & RA & Dec & & ID & RA & Dec \\
\hline
178 & 28.5414 & -44.033 & & 179 & 20.7624 & -43.127 & & 180 & 18.0030 & -33.564 \\
181 & 19.9193 & -33.100 & & 182 & 13.5783 & -23.554 & & 183 & 17.9328 & -23.305 \\
184 & 18.9117 & -23.255 & & 185 & 20.1795 & -14.732 & & 186 & 28.5135 & -14.253 \\
187 & 41.2831 & -13.236 & & 188 & 27.6655 & -3.9382 & & 189 & 13.5655 & -3.5895 \\
190 & 18.2573 & -64.826 & & 191 & 13.9032 & -63.736 & & 192 & 29.5756 & -54.215 \\
193 & 14.0309 & -53.189 & & 194 & 14.2397 & -52.923 & &  & &  \\
\hline
\end{tabular}
\caption{The coordinates of detected possible galaxies with dust lanes.}
\label{dust_lane}
}
\end{table*}

\begin{table*}[h!t]
{
\centering
\begin{tabular}{|l|c|c|c|c|c|c|c|c|c|c|}
\hline
ID & RA & Dec  & & ID & RA & Dec & & ID & RA & Dec \\
\hline
195 & 25.0351 & -43.740 & & 196 & 46.5785 & -39.344 & & 197 & 71.0259 & -38.484 \\
198 & 66.6586 & -36.038 & & 199 & 91.7963 & -35.067 & & 200 & 19.9831 & -34.265 \\
201 & 11.0489 & -24.322 & & 202 & 10.7240 & -23.545 & & 203 & 13.9377 & -23.478 \\
204 & 42.8310 & -13.355 & & 205 & 1.49017 & -4.9928 & & 206 & 0.18903 & -4.7813 \\
207 & 7.08776 & -4.8255 & & 208 & 14.8486 & -4.8032 & & 209 & 11.2310 & -4.1358 \\
210 & 1.02484 & -4.0871 & & 211 & 24.3611 & -64.896 & & 212 & 15.9150 & -2.3332 \\
213 & 1.37269 & -2.0939 & & 214 & 0.32218 & -1.5197 & & 215 & 11.4851 & -63.729 \\
216 & 16.1720 & -63.829 & & 217 & 33.0961 & -60.101 & & 218 & 230.280 & 50.6720 \\
219 & 26.1203 & -49.790 & & 220 & 22.4344 & -48.891 & & 221 & 13.2715 & -2.4701 \\
222 & 358.047 & 11.4678 & &  & &    & &  & &  \\
\hline
\end{tabular}
\caption{The coordinates of detected objects that can be tidally distorted systems.}
\label{tidially_distorted}
}
\end{table*}

\begin{table*}[h!t]
{
\centering
\begin{tabular}{|l|c|c|c|c|c|c|c|c|c|c|}
\hline
ID & RA & Dec  & & ID & RA & Dec & & ID & RA & Dec \\
\hline
223 & 14.2833 & -43.727 & & 224 & 23.8920 & -43.410 & & 225 & 54.8327 & -38.975 \\
226 & 77.3550 & -38.464 & & 227 & 6.01508 & -34.960 & & 228 & 12.9062 & -34.886 \\
229 & 23.4023 & -29.885 & & 230 & 23.7417 & -29.759 & & 231 & 22.3487 & -28.631 \\
232 & 11.7824 & -24.376 & & 233 & 10.7661 & -22.247 & & 234 & 10.7245 & -22.257 \\
235 & 25.2616 & -14.314 & & 236 & 23.2281 & -13.864 & & 237 & 42.8663 & -13.298 \\
238 & 28.5265 & -4.8778 & & 239 & 5.04152 & -4.4164 & & 240 & 11.0260 & -4.4902 \\
241 & 17.8263 & -4.0600 & & 242 & 22.3028 & -3.8874 & & 243 & 11.9790 & -3.5191 \\
244 & 12.1786 & -64.994 & & 245 & 5.06276 & -2.6528 & & 246 & 4.75718 & -2.4539 \\
247 & 16.0499 & -2.0576 & & 248 & 7.40026 & -1.7385 & & 249 & 122.447 & 5.01916 \\
250 & 33.2834 & -61.752 & & 251 & 36.5613 & -61.631 & & 252 & 30.8603 & -61.218 \\
\hline
\end{tabular}
\caption{The coordinates of other galaxies identified as outlier galaxies.}
\label{others}
}
\end{table*}

\subsection{Performance evaluation}

One of the considerations of the described algorithm is response time. The bottleneck of the analysis is the representation of each image by a set of its numerical content descriptors. Analysis of a single image required nearly two minutes using a single core of an Intel Core-i7 processor. That means that a single core can analyze the entire set of $\sim2\cdot10^6$ galaxies in nearly eight years. 

To handle the data, 32 cores of a Beowulf cluster were used, and reduced the response time of the system to $\sim3$ months of computing. While digital sky surveys are becoming increasingly more powerful, computing resources, and especially parallelizations, are also becoming more accessible. For instance, modern processors have 64 or more cores, and the availability of multiple cores in single processors is expected to grow. Therefore, while the method is computationally demanding, its requirement of computing power can be matched by the increasing availability of hardware that can be parallelized. Specifically, graphics processing units (GPUs) can be customized to parallelize the analysis, and perform a faster analysis by reducing the energy and cost of the hardware.

\section{Conclusions}
\label{conclusion}

Autonomous digital sky surveys can acquire very large databases of astronomical data, making ``traditional" manual analysis of the data impractical. 
Perhaps one of the more algorithmically challenging tasks is identification of peculiar astronomical object of potential interest among millions of other common astronomical objects. 

This study applies a method of automatic detection of outlier galaxies imaged by the Dark Energy Survey. The experiment shows that an automated method can provide a practical solution to the problem of identification of peculiar galaxies in large databases. Although the automatic identification results in a large number of false positives, it allows to reduce the size of the data that needs to be inspected to make manual detection practical. That allows to reduce a dataset of millions of objects into a far smaller dataset of thousands of objects, of which several hundred objects are outliers.

As automatic detection of peculiar galaxies is a relatively complex task, the method shown in this paper is clearly not perfect. The advantage of the method is that it allows to control the number of alerts, and consequently handle very large image databases. The initial dataset used here of $2\cdot10^6$ objects was reduced to 3$\cdot 10^3$ objects, where $\sim92$\% of these objects were not outlier galaxies. That makes a false-positive rate of $1.375 \cdot 10^{-3}$. The true positive is obviously much lower, at $1.125 \cdot 10^{-4}$.

Digital sky surveys have been growing consistently in both power and number, and that trend in astronomy research is bound to continue. Surveys such as Vera Rubin Observatory and the space-based Euclid are further expanding the already high throughput of modern ground-based and space-based sky surveys. Due to the size of the data, it can be reasonably assumed that many objects of paramount scientific interest will be hidden in these databases, but might not be noticed. The method described here can be easily applied to data from the fourth-generation surveys and provide a large number of irregulars galaxies, peculiar galaxies, strong gravitational lenses, etc. This will significantly benefit studies of galaxy evolution and the exploration of cosmologies.








\section*{Acknowledgment}

I would like to thank the anonymous reviewer for providing comments that improved the manuscript. The research was funded by NSF grant AST-1903823.




\end{document}